\def\Tr{\mathop{\rm Tr}\nolimits} 
\def\bra#1{\langle #1 \,\vert}
\def\ket#1{\vert\, #1 \rangle}
\def\frac#1#2{{#1 \over #2}}
\def\e{\mbox{e}}
\def\d{\mbox{d}}
\def\Dirac#1{#1\hskip-6pt/}
\def\dd{\Dirac\partial}
\documentstyle[prl,aps]{revtex}
\begin{document}
\title{NJL model with vector couplings vs. phenomenology}
\author{Chr.V.~Christov\thanks{
Permanent address:Institute for Nuclear Research and Nuclear
Energy, Sofia, Bulgaria}
\thanks{christov@neutron.tp2.ruhr-uni-bochum.de},
K.~Goeke\thanks{goeke@hadron.tp2.ruhr-uni-bochum.de}}
\address{Institut f\"ur Theoretische Physik II,
Ruhr-Universit\"at Bochum, D-44780 Bochum, Germany}
\author{M.~Polyakov\thanks{maxpol@lnpi.spb.su}}
\address{Petersburg Nuclear Physics Institute, Gatchina,
St.Petersburg 188350, Russia}
\maketitle

\begin{abstract}
We study the pseudoscalar, vector and axial current correlation functions
in SU(2)-NJL model with scalar and vector couplings. The correlation
functions are evaluated in leading order in number of colors $N_c$. As it is
expected in the pseudoscalar channel pions appear as Goldstone bosons, and
after fixing the cutoff to reproduce the physical pion decay constant, we
obtain well-known current-algebra results. For the vector and axial channels
we use essentially that at spacelike momenta the correlation functions
can be related to the experimentally known spectral density via
dispersion relations. We show that the latter imposes strong
bounds on the strength of the vector coupling in the model. We find that
the commonly used on-shell treatment of the vector and axial mesons
(identified as poles at large timelike momenta) fails to reproduce the
behavior of the corresponding correlation functions at small spacelike
momenta extracted from the physical spectral density. The parameters of the
NJL model fixed by the correlation functions at small spacelike momenta
differ noticeably from those of the on-shell treatment.
\end{abstract}
\draft
\pacs{PACS number(s): 12.39.Fe,11.15.Pg,11.55.Fv}

\vspace {-10.5cm}
\baselineskip=0.4cm
\begin{flushright}{RUB-TPII-19/94\\December, 94}
\end{flushright}
\vspace {10.5cm}

The correlation functions of different colorless hadronic currents built of
quarks and gluons offer a general tool to study the structure of the QCD
vacuum and the hadron spectrum~\cite{Shuryak93}. One can obtain information
about the
physical intermediate states from the pole structure of the correlation
functions in the timelike region as well as from its behavior at
spacelike momenta.
For an exact theory both ways should lead to identical results.
This, however, is not necessarily true for the effective models, which
in the present lack of non-perturbative solution to QCD  attract an
increasing interest. Among them, the Nambu -- Jona-Lasinio model,
proposed~\cite{Nambu61} in early sixties
in analogy with the BCS theory of superconductivity, plays an
essential role. This model incorporates the basic
symmetries of QCD and in particular, the chiral symmetry. It is able to
manifest the spontaneous chiral symmetry breaking and, as a  natural
consequence, the appearance of the pseudoscalar mesons as Goldstone
particles. As such it offers a simple but
quite successful working scheme to study the role of the dynamical chiral
symmetry breaking mechanism in hadron physics (for review see
refs.~\cite{Weise92,Volkov93,Hatsuda94,Meissner94}). However, the model
suffers from two important drawbacks. First, it is not renormalizable
and to make the theory finite one needs a finite cut off fixed in a
consisitent way (usually to reproduce the pion decay constant) and
the limit of large $N_c$ (number of colors).
The typical value of the cut off is about 1 GeV. The second problem is the
lack of confinement. Obviously, these drawbacks make the so far used
on-shell treatment of the (axial) vector
mesons questionable~\cite{Bernard88,Klimt90,Blin90,Ripka92,Bijnens94}.
In the present work we will show that the proper way to overcome this
problem is to consider the corresponding current correlation function at
small spacelike momenta. In this case we do not face the problem
of the strong coupling to the $\bar qq$ continuum and we apply the
model at small momentum region where it is designed for. To this end we make
essentially use of the fact that the correlator at small  spacelike
momenta can be related via dispersion relations to well-established
rich phenomenology without detailed knowledge of
the dynamics at large timelike momenta. We will show that such a scheme
allows to describe the vector modes in the meson sector of the
NJL model in a consistent way.

The simplest SU(2) Nambu--Jona-Lasinio (NJL) model is defined by
lagrangean~\cite{Nambu61}:
\begin{eqnarray}
{\cal L}_{NJL} = \bar \Psi \, (i \dd - m_0)\, \Psi &+& {G_s\over 2} \,
[(\bar \Psi \Psi )^2 + (\bar \Psi i \vec\tau \gamma_5 \Psi )^2\,] \nonumber\\
&-&{G_v\over 2} \, [(\bar \Psi\gamma_\mu \vec\tau \Psi )^2 + (\bar \Psi
\vec\tau\gamma_\mu \gamma_5 \Psi )^2\,] \,,\label{NJLLAG}
\end{eqnarray}
which contains chirally invariant local four-fermion interactions with scalar
$G_s$ and vector $G_v$ couplings\footnote{This lagrangean with vector
couplings was considered first by Kikkawa~\cite{Kikkawa76}}.The current
quark mass $m_0$
stands for both $up$ and $down$ quarks (assumed to be degenerated in mass).

Applying the well-known
bosonization procedure~\cite{Eguchi74} we arrive at
the model generating functional expressed as a functional integral over
auxiliary boson fields $S, {\vec P}, {\vec V_\mu}, {\vec A_\mu} $:
\begin{equation}
Z_{NJL} = \int{\cal D}S{\cal D}\vec P {\cal D}\vec V_\mu{\cal
D}\vec A_\mu\, \e^{N_c\Tr\log\Bigl(D[S,{\vec P},{\vec V_\mu},{\vec
A_\mu}]\Bigr)+ I_m[S,{\vec P},{\vec V_\mu},{\vec A_\mu}]}\,,
\label{GENF} \end{equation}
where the quarks are already integrated out. To simplify our notations we
adopt for $P$, $V_\mu$ and $A_\mu$ that $P\equiv\vec P\cdot\vec\tau$ etc.
The euclidean Dirac operator $D$ is given by
\begin{equation}
D=  - i \partial_\mu\gamma_\mu+
S + i\, P \gamma_5 + V_\mu\gamma_\mu + A_\mu\gamma_5 \gamma_\mu
\label{DIRAC} \end{equation}
and the trace over colors is explicitly done. The symmetry breaking
term is included in the boson part:
\begin{equation}
I_m= \int\d^4x \Bigl\{{1\over 2G_s} (S^2 + {\vec P}^2)+{1\over 2G_v}
({\vec V_\mu}^2 + {\vec A_\mu}^2)-{m_0\over G_s}S\Bigr\}\,.
\label{MESON} \end{equation}

In the large $N_c$ limit the integral in eq.(\ref{GENF}) is given
by its saddle point value. For the translational invariant case (vacuum) we
have a non-trivial stationary meson configuration given by
\begin{equation}
S=M, \quad \vec P=0, \quad\vec V_\mu=0, \quad\vec A_\mu=0\,,
\label{VACUUM} \end{equation}
which means that the chiral symmetry is spontaneously broken. The quarks
acquire a constituent mass $M$ which is related to the scalar coupling
constant $G_s$ via the stationary condition (saddle point):
\begin{equation}
M=2G_sMN_c\Tr\Bigl({1\over D[M]}\Bigr)+m_0=m_0-G_s< \bar \Psi \Psi >\,.
\label{GAP} \end{equation}
Here $< \bar \Psi \Psi >$ is the chiral quark condensate.
As it was mentioned the model is not renormalizable and in order to make the
fermion determinant finite it must be regularized.
Thus, the model has four parameters,
namely two coupling constants $G_s$ and $G_v$, the current mass $m_0$ and
the cut off $\Lambda$. Usually they are fixed reproducing the physical pion
decay constant and the physical masses of pion and rho meson. It leaves
one parameter
free which is commonly chosen to be the constituent mass $M$ because with
eq.(\ref{GAP}) one can eliminate $G_s$ in favor of $M$. In principle, one
can use the empirical value for the quark condensate $< \bar \Psi \Psi >$ to
constrain $M$. However, being quadratically divergent $< \bar \Psi \Psi >$
is very sensitive to the details of the regularization scheme.

In this paper we suggest an alternative way of fixing the model
parameters $G_s$, $G_v$, $m_0$ and $\Lambda$. To this end we study the
correlation functions of the pseudoscalar, vector and axial-vector currents:
\begin{equation}
P^a(x)=\bar\Psi(x)i\gamma_5{\tau^a\over 2}\Psi\,,
\label{PCURR} \end{equation}
\begin{equation}
V^a_\mu(x)=\bar\Psi(x)\gamma_\mu{\tau^a\over 2}\Psi\,,
\label{VCURR} \end{equation}
\begin{equation}
A^a_\mu(x)=\bar\Psi(x)\gamma_\mu\gamma_5{\tau^a\over 2}\Psi\,,
\label{ACURR} \end{equation}
in the NJL model.
The correlation functions are defined by the vacuum-to-vacuum matrix
elements of the currents and can be expressed in terms of invariant
functions as follows: \begin{equation}
\Pi_P(Q^2)= \int \d_4x\e^{iqx}\bra{0}{\cal
T}\{P^a(x)P^a(0)\}\ket{0}\,,
\label{PCRDEFF} \end{equation}
\begin{equation}
(q_\mu q_\nu-q^2g_{\mu\nu})\Pi_v(Q^2)= \int \d_4x\e^{iqx}\bra{0}{\cal
T}\{V^a_\mu(x)V^a_\mu(0)\}\ket{0}\,,
\label{VCRDEFF} \end{equation}
\begin{equation}
(q_\mu q_\nu-q^2g_{\mu\nu})\Pi_a^{t}(Q^2)+q_\mu q_\nu\Pi_a^l(Q^2)=
\int \d_4x\e^{iqx}\bra{0}{\cal T}\{A^a_\mu(x)A^a_\mu(0)\}\ket{0}\,,
\label{ACRDEFF} \end{equation}
where as usual $Q$ is defined as $Q^2=-q^2$. In contrast to the vector
current in the real world the axial-vector current is not conserved and
the correlation function in the axial channel has
an additional longitudinal part $\Pi_a^l(Q^2)$ which vanishes in the chiral
limit $m_0\rightarrow 0$. In the NJL model\footnote{Using  different
techniques the invariant functions are evaluated also in
ref.~\cite{Bijnens94}} the matrix element of rhs
of eqs.(\ref{PCRDEFF}-\ref{ACRDEFF}) can be expressed as a path integral
using the model generating functional (\ref{GENF}). In the leading order in
$N_c$ only two diagrams (shown in fig.~\ref{Figr1}) contribute to the matrix
elements. The first one is a simple one-quark loop whereas the second
includes a boson line.
For the pseudoscalar function we get the following result:

\begin{equation}
\Pi_a(Q^2)={Q^2Z_p(Q^2)f_a(Q^2)-{< \bar \Psi \Psi >\over M}\over
G_sf_a(Q^2)Z_p(Q^2)}{1\over Q^2+{m_0\over M G_s}{1\over
B_p(Q^2)f_a(Q^2)}}\,, \label{PQ}
\end{equation}
where
\begin{equation}
f_a(Q^2)={1\over 1+4M^2G_vZ_p(Q^2)}\,.
\label{GA} \end{equation}
Function $Z_p(Q^2)$ corresponds to the quark-loop diagram with two
pseudoscalar-isovector vertices ($i\gamma_5\tau_a$).
As already mentioned, the model is not renormalizable and in order to make
the quark loop finite we need a regularization. In our calculations we
use the proper-time as well as the Pauli-Villars regularization scheme.
Both schemes preserve the symmetries of the model.
Here we present only the proper-time regularized expression for
$Z_p(Q^2)$~\cite{Ripka92}:
\begin{equation}
Z_p(Q^2)={4 N_C \over (2\pi)^4}\int\limits_{-1}^{+1}{\d u\over
2}\int\limits_{\Lambda^{-2}}^\infty {\d s\over s}\e^{-s\biggl[M^2+{Q^2\over
4}(1-u^2)\biggr]}\,,
\label{Bpq}
\end{equation}
and for the condensate:
\begin{equation}
< \bar \Psi \Psi >=-{N_C \over 2\pi^2}M \int\limits_{\Lambda^{-2}}^\infty
{\d s\over s^2}\e^{-s M^2}\,.
\label{QQ}
\end{equation}
Details as well as regularized expressions for the quark condensate and the
function $Z_p(Q^2)$ in the case of Pauli-Villars regularization can be found
in ref.~\cite{Schuren93}.

In the chiral limit the pseudoscalar correlation function
(\ref{PQ}) develops the expected Goldstone pole at $Q^2=0$:
\begin{equation}
\Pi_a(Q^2\rightarrow 0)\longrightarrow {< \bar \Psi \Psi >^2\over Q^2
f_\pi^2}
\label{PQ0}
\end{equation}
where the pion decay constant $f_\pi^2$ is given by the residue of the
pole:
\begin{equation}
f_\pi^2=M^2f_a(0)Z_p(0)\,.
\label{Fpi} \end{equation}

The non-zero current mass $m_0$ shifts the pion pole in
eq.(\ref{PQ0}) to $Q^2=-m^2_\pi$ and one recovers
the Gell-Mann - Oakes - Renner relation:
\begin{equation}
f_\pi^2m^2_\pi=-m_0<\bar \Psi \Psi>+O(m_0^2)\,.
\label{GMOR} \end{equation}

In the vector channel the correlation function $\Pi_v(Q^2)$ is given by
\begin{equation}
\Pi_v(Q^2)={1\over 4}\,{Z_v(Q^2)\over 1+G_vZ_v(Q^2)Q^2}\,,
\label{PVQ2} \end{equation}
where, similarly to $Z_p(Q^2)$, the function  $Z_v(Q^2)$ corresponds to the
quark-loop diagram with two vector vertices ($\gamma_\mu\tau^a$) (lhs of
fig.~\ref{Figr1}). In particular, in the proper-time scheme it has the
following form:
\begin{equation}
Z_v(Q^2)={4 N_C \over (2\pi)^4}\int\limits_{-1}^{+1}{\d u\over
2}(1-u^2) \int\limits_{\Lambda^{-2}}^\infty {\d s\over
s}\e^{-s\biggl[M^2+{Q^2\over 4}(1-u^2)\biggr]}\,.
\label{Bvq}
\end{equation}

For the transverse invariant function of the axial current correlator in the
NJL model we have:
\begin{equation}
Q^2\Pi^t_a(Q^2)={1\over 4}{Z_v(Q^2)Q^2+4M^2Z_p(Q^2)\over
1+G_vZ_v(Q^2)Q^2+4G_vM^2Z_p(Q^2)}\,,
\label{PTAQ0} \end{equation}
whereas the longitudinal one includes only a pion contribution:
\begin{equation}
Q^2\Pi^l_a(Q^2)=-{f_\pi^2m_\pi^2 \over
Q^2+{m_\pi^2f_\pi^2\over M^2Z_p(Q^2)f_a(Q^2)}}\,.
\label{PLAQ} \end{equation}

The correlation functions (\ref{PVQ2}),(\ref{PTAQ0}) possess (at some
model parameter values) poles in the timelike region $Q^2<0$ which is in
fact the pole of the propagator of the auxiliary (axial) vector field. This
is used in the so-called on-shell
treatment \cite{Bernard88,Klimt90,Blin90,Takizawa91,Ripka92,Bijnens94}
where one fixes the vector coupling
constant $G_v$ adjusting the position of the pole to the physical rho mass.
The other model parameters, namely the cutoff $\Lambda$ and the current mass
$m_0$ are fixed reproducing the physical pion mass and the physical pion
decay constant from eqs.~(\ref{Fpi}),(\ref{GMOR}), whereas
the constituent quark mass is treated as a free parameter as usual.
To illustrate the on-shell scheme we present in fig.\ref{Figr2} the
corresponding parameter values (index ``pole'') as a function of $M$
for the case of the the Pauli-Villars regularization. Since the results
with the proper-time regularization
are almost identical with those of the proper-time ones we do not discuss
them
separately. As can be seen the behavior of the coupling constants and the
cutoff as function of the constituent mass is not smooth -- the
value $M=m_\rho^{exp}/2$ is a singular point where all curves show a kink.
At  $M > m_\rho^{exp}/2$ the vector current correlation function has a real
pole
at $Q^2=-m_\rho^2$ which is an indication for a bound state in the spectrum.
However, because of lack of confinement a second peak in the
$\bar qq$ continuum appears, which is not taken into account in the
definition of the physical $\rho$ field. Obviously, it is hard to relate the
second peak to the excited $\rho(1450)$-resonance. At smaller values of
$M<m_\rho^{exp}/2$ in the vector channel a broad resonance appears in
the $\bar qq$ continuum because the corresponding pole moves to the complex
plane of $Q^2$. The position of the peak depends on the vector coupling
constant $G_v$. Using this\footnote{Such a procedure was used in
refs.\cite{Klimt90,Blin90}} we determine $G_v$ fixing
the position of the peak in the vector spectral function $\mbox{Im}\Pi_v$
at the physical $\rho$ mass.  However, as can be seen from
fig.\ref{Figr2} it is not always possible to find a solution for $G_v$: for
$M$ between 230 and 260 MeV in fact there is no such a solution.
It should be also noted that at $M > m_\rho^{exp}/2$ both $G_v$
and $G_s$, as well as the cutoff $\Lambda$ show a stronger dependence on $M$
as $G_v$ is dominant ($G_v > G_s)$. This changes for smaller values
of $M$ where $G_s$ and $\Lambda$ stay almost constant whereas $G_v$
decreases. In the
axial channel for the values of the constituent mass $M$ considered (200 -
450 MeV) the $A_1$ meson appears as a very broad peak being centered around
1 GeV in the $\bar qq$ continuum. In Fig.\ref{Figr2} we also show the
parameter values (index ``grad'') obtained by means of a
gradient or heat-kernel expansion of the effective
action~\cite{Ebert83,Ebert86}. This procedure is frequently used
to fix the vector coupling constant $G_v$ in
NJL model and in fact it is an approximation to the on-shell description of
vector modes. As can be seen from Fig.\ref{Figr2} this procedure provides
a quite crude approximation to the on-shell values because of the large
masses of the vector mesons. From the above discussion one concludes
that the on-shell treatment is not able to provide a consistent description
of the vector meson modes in NJL model.

As a next step we suggest to
make use of the fact that at spacelike momenta $Q^2>0$ the
vector and axial-vector current correlation functions can be related to the
experimentally known spectral density via dispersion relations. In the
vector channel the invariant function $\Pi_v(Q^2)$ satisfies a dispersion
relation with one subtraction:
\begin{equation}
\Pi_v(Q^2)=\Pi_v(0)-{Q^2\over \pi}\int\d s { \mbox{Im}\Pi_v(s)\over
s(s+Q^2)}\,, \label{DISPREL} \end{equation}
where the physical spectral density $\mbox{Im}\Pi_v(s)$ is experimentally
accessible:
\begin{equation}
\mbox{Im}\Pi_v(s)={1\over 12\pi}{\sigma_{e^+e^- \rightarrow
h}(s)\over\sigma_{e^+e^- \rightarrow \mu^+\mu^-}(s)}\,.
\label{SPECTR} \end{equation}
For the rhs we use a simple ($\rho$ + continuum)
parameterization~\cite{Shifman79} of the experimental data
\begin{equation}
\mbox{Im}\Pi_v(s)={\pi m^2_\rho\over g^2_\rho}\delta(s-m^2_\rho)+{1\over
8\pi}\Bigl(1+{\alpha_s\over \pi}\Bigr)\Theta(s-s_0)\,,
\label{EXPSPV} \end{equation}
with the parameter values:
\begin{equation}
m^2_\rho=0.77\, \hbox{GeV},\quad \Bigl({g^2_\rho\over 4\pi}\Bigr)=2.36,
\quad s_0=1.5\, \hbox{GeV}^2\,.
\label{VPARAM} \end{equation}

Eqs.(\ref{DISPREL}) and (\ref{EXPSPV}) provide an independent way to fix
the vector coupling constant $G_v$ confronting the model with
the experiment by equating the phenomenological dispersion integral and the
model correlation function at low positive $Q^2$:
\begin{equation}
\Pi_v(0)-{Q^2\over \pi}\int\d s { \mbox{Im}\Pi_v(s)\over
s(s+Q^2)}={1\over 4}{Z_v(Q^2)\over
1+G_vZ_v(Q^2)Q^2}\,. \label{PhVQ2} \end{equation}

The resulting $G_v$ as well as $\Lambda$ and $G_s$ as a function of $M$,
obtained by adjusting ($\chi^2$-fit) the lhs of eq.~(\ref{PhVQ2}) to the
rhs with the physical spectral density (\ref{EXPSPV}) at low positive
momenta (we consider $0<Q^2<0.5$ GeV$^2$), are shown (index ``corr'') in
fig.\ref{Figr2} a) and b). Apparently the correlation
function at spacelike momenta and the on-shell treatment yield rather
different results. In contrast to the curves resulting from the on-shell
scheme those of the correlation function have the advantage to behave
smoothly as a function of $M$. This, and the very fact that the NJL model is
designed for low-energy structures involving small momenta, makes the above
correlation method preferable. With increasing $M$ larger values for $G_v$
are needed to fit the phenomenological side whereas the $G_s$ is slightly
decreasing.

In the axial channel, since the longitudinal part of the axial correlation
function is almost independent on the $G_v$ and $M$, we will concentrate our
further discussion on its transverse part $\Pi_a^{t}$. It
obeys a dispersion relation with two subtractions:
\begin{equation}
Q^2\Pi_a^{t}(Q^2)=\Pi_a^{t}(0)+Q^2{\d \Pi_a^{t}(0)\over \d
Q^2}+{q^4\over \pi}\int \d s {\mbox{Im} \Pi_a^{t}(s)\over s^2(s+Q^2)}\,.
\label{PAT}
\end{equation}
Similar to the vector channel for
$\mbox{Im}\Pi_a^{t}(Q^2)$ we take a simplified ($A_1$ + continuum)
parameterization~\cite{Shifman79}
\begin{equation}
\mbox{Im}\Pi_a^{t}(s)={\pi m^4_{A_1}\over g^2_\rho}\delta(s-m^2_{A_1})+{1\over
8\pi}\Bigl(1+{\alpha_s\over \pi}\Bigr)\Theta(s-s_0)\,,
\label{EXPSPA} \end{equation}
with parameters~\cite{Reinders82}:
\begin{equation}
m^2_{A_1}=1.26\, \hbox{GeV},\quad \Bigl({4\pi\over g^2_{A_1}}\Bigr)=0.15 -
0.18, \quad s_0=1.7\, \hbox{GeV}^2\,.
\label{APARAM} \end{equation}

Confronting the model
$A_1$-correlation function~(\ref{PTAQ0}) and the
phenomenological expressions~(\ref{PAT}),~(\ref{EXPSPA}) one
has another way to fix the vector coupling constant
$G_v$. The obtained vector coupling constant ($G_{v(a_1)}^{corr}$) as a
function of $M$ is shown in fig.\ref{Figr2} a)
in comparison with the results from fitting the $\rho$ phenomenology as well
as from the on-shell $\rho$ treatment and the derivative expansion. In
contrast to the vector channel the model axial function $\Pi_a^{t}(Q^2)$
depends much stronger on the constituent mass and because of that the
model is able to fit simultaneously both $\rho$- and $A_1$-channels
only within a narrow window of values for the constituent quark mass $M$
around 240 MeV. It is interesting to note that in this window the on-shell
values are also very close to those obtained from the correlation functions.
The particular value of $M$ depends on the parameter
values (\ref{VPARAM}),(\ref{APARAM}) used for the (resonance + continuum)
parameterizations. In fact the estimate of the constituent mass $M\approx
300$ MeV in the random instanton model~\cite{Shuryak88,Pobylitsa89}
of QCD vacuum, applied successfully also to the meson correlation
functions~\cite{Shuryak93a}, is not far from our numbers. The
resulting vector coupling constant $G_v$ is of the same order
as the scalar one $G_s$. In this point we disagree with
ref.~\cite{Zakharov94} in which constraining the NJL
model via QCD sum rules it is concluded that $G_v$ should be almost zero or
at least $G_v\ll G_s$. It is due to the fact that in contrast to us in
ref.~\cite{Zakharov94} the NJL model is treated in a quite simplified
approximate way ignoring the full $Q^2$-dependence of the correlation
functions as well as the relations between the model parameters fixed to
reproduce the physical pion properties.

In our calculations so far we used simple parameterizations
(\ref{EXPSPV}),(\ref{EXPSPA}) for the experimental spectral densities in
which the resonance
widths are neglected. In order to check this approximation we repeated the
calculations using a much more elaborated (finite-width resonance+continuum)
parameterization~\cite{Shuryak94} of the experimental data. The obtained
results follow qualitatively the picture of fig.~\ref{Figr2}. The only
difference is that the narrow window, where in both vector and axial
channels the experimental low-energy behavior of the correlation functions
can be reproduced simultaneously, is shifted to higher
values of the constituent mass $M\approx 260$ MeV.

We also related the model with $G_v$, fixed in the present scheme,
to the chiral effective lagrangean of Gasser and
Leutwyler~\cite{Gasser85}. Following
ref.~\cite{Arriola91} we calculated the corresponding low-energy
coefficients $\bar l_1 - \bar l_6$ in the notation of
ref.~\cite{Gasser85} and low-energy pionic characteristics. As can
be seen from Table~\ref{Tabl1} our theoretical predictions are in good
agreement with the empirical values.

To summarize, in this letter we study the vector meson modes in the NJL
model in terms of current correlation functions. We find that the on-shell
treatment of the vector meson fails to reproduce the low-energy behavior of
the correlation functions fixed by the experimentally known spectral density
via dispersion relations. We offer a different scheme to fix the NJL-model
parameters based on the behavior of the correlation functions at small
spacelike momenta. Treating
the vector coupling constant as a free parameter the model is able to
reproduce the phenomenology in both the axial- and vector channels for the
constituent mass only in a narrow window around 240 MeV. The vector
coupling constant is by no means zero and is of the same order as
the scalar coupling constant. In principle, these two schemes, the on-shell
treatment (large timelike momenta) and the one based on the behavior of the
correlation function
at small spacelike momenta, should provide identical results within an
``exact'' theory, which however is not the case for an effective model like
NJL. In the latter case the correlation function method appears to be
preferable. One
also should keep in mind that the present considerations are done in leading
order in $N_c$. The inclusion of $1/N_c$ quantum boson (loop)
corrections could in principle change the present results.

\section*{Acknowledgement}

We would like to thank Dmitri Diakonov, Victor Petrov and Georges Ripka for
helpful discussions. The project has been partially supported
by the VW Stiftung, DFG and COSY (J\"ulich).

\newpage
\begin{figure}
\caption{Diagrams contributing to the correlation functions in leading $N_c$
order.}
\label{Figr1}
\end{figure}
\begin{figure}
\caption{The parameters of the model as
function of the constituent mass $M$:
a) vector coupling constant $G_v$  fixed in the on-shell
treatment  $G_v^{pole}$, from derivative expansion $G_v^{grad}$, and from
the correlation functions in the case
of $\rho$- and $a_1$-phenomenology fits; b) the same as a) but for the
scalar coupling constant $G_s$ and the cut off $\Lambda$. }
\label{Figr2}
\end{figure}

\newpage
\begin{table}
\caption{Low-energy SU(2) coefficients and low-energy pion
properties for constituent mass $M=235$ MeV.}
\label{Tabl1}
\begin{tabular}{cccccc}
SU(2) coefficients & model & phenom. & $\pi$ properties  & model & exp. \\
\hline $\Lambda$ [MeV] & 1400  &      & $a^0_0$ & 0.16 & $0.26\pm 0.05$ \\
$-<\bar qq>^{1/3}$ [MeV] & 310  & $265\pm 40$ & $b^0_0$ & 0.19 & $0.25\pm
0.03$ \\ 
$m_0$ [MeV] & 3.5 & $7\pm 3$ & $a^2_0$ & $-0.045$ &$ -0.028\pm 0.012$ \\
$\bar l_1$ &$-4.17$ & $-4.56\pm 1.5$  & $b^2_0$ & $-0.084$ &$ -0.082\pm
0.008$ \\ 
$\bar l_2$ & 3.02 &$ 2.3\pm 0.7$  & $a^1_1$ & 0.035 &$ 0.038\pm 0.002$ \\
$\bar l_3$ & 2.45 &$-1.0\pm 3$   & $a^0_2\times 10^{4}$ & 9.8 &$ 17.0\pm
3.0$ \\ 
$\bar l_4$ & 2.44 &$1.1\pm 0.4$  & $a^2_2\times 10^{4}$ & $-1.1$ &$ 1.0\pm
3.0$ \\ 
$\bar l_5$ & 9.1  & $10.4\pm 1.3$  & $<r^2>_s^\pi [\mbox{fm}^2$] & 0.43 &
$0.52\pm 0.15$ \\ 
$\bar l_6$ & 12.7 & $13.1\pm 1.3$  & $<r^2>_v^\pi [\mbox{fm}^2$] & 0.37 &$
0.44\pm 0.03$ \\ 
\end{tabular}

\end{table}

\end{document}